\documentclass[aps,prb,reprint,showpacs,showkeys,superscriptaddress]{revtex4-1} 

\usepackage{amssymb,amsmath,graphicx,natbib,hyperref}
\begin{document}
\title{Acoustically regulated carrier injection into a single optically active quantum dot}

\author{Florian J. R. Sch\"{u}lein}
\affiliation{Lehrstuhl f\"{u}r Experimentalphysik 1 and Augsburg Centre for Innovative Technologies ({\it ACIT}), Universit\"{a}t Augsburg, Universit\"{a}tsstr. 1, 86159 Augsburg, Germany} 
\affiliation{Center for NanoScience ({\it CeNS}), Ludwig-Maximilians-Universit\"at M\"unchen, Geschwister-Scholl-Platz 1, 80539 M\"{u}nchen, Germany}
\author{Kai M\"{u}ller}
\author{Max Bichler}
\author{Gregor Koblm\"uller}
\author{Jonathan J. Finley}
\affiliation{Walter Schottky Institut and Physik Department, Technische Universit\"at M\"unchen, Am Coulombwall 4a, 85748 Garching, Germany}
\author{Achim Wixforth}
\affiliation{Lehrstuhl f\"{u}r Experimentalphysik 1 and Augsburg Centre for Innovative Technologies ({\it ACIT}), Universit\"{a}t Augsburg, Universit\"{a}tsstr. 1, 86159 Augsburg, Germany} 
\affiliation{Center for NanoScience ({\it CeNS}), Ludwig-Maximilians-Universit\"at M\"unchen, Geschwister-Scholl-Platz 1, 80539 M\"{u}nchen, Germany}
\author{Hubert J. Krenner}\email{hubert.krenner@physik.uni-augsburg.de}
\affiliation{Lehrstuhl f\"{u}r Experimentalphysik 1 and Augsburg Centre for Innovative Technologies ({\it ACIT}), Universit\"{a}t Augsburg, Universit\"{a}tsstr. 1, 86159 Augsburg, Germany} 
\affiliation{Center for NanoScience ({\it CeNS}), Ludwig-Maximilians-Universit\"at M\"unchen, Geschwister-Scholl-Platz 1, 80539 M\"{u}nchen, Germany}

\preprint{}

\begin{abstract}
We study the carrier injection into a single InGaAs/GaAs quantum dot regulated by a radio frequency surface acoustic wave. We find that the time of laser excitation during the acoustic cycle programs both the emission intensities and time of formation of neutral $(X^0)$ and negatively charged $(X^-)$ excitons. We identify underlying, characteristic formation pathways of both few-particle states in the time-domain experiments and show that both exciton species can be formed either with the optical pump or at later times by injection of single electrons and holes "surfing" the acoustic wave. All experimental observations are in excellent agreement with calculated electron and hole trajectories in the plane of the two-dimensional wetting layer which is dynamically modulated by the acoustically induced piezoelectric potentials. Taken together, our findings provide insight on both the onset of acousto-electric transport of electrons and holes and their conversion into the optical domain after regulated injection into a single quantum dot emitter.
\end{abstract}

\maketitle
\section{Introduction}
Deterministic transport of individual charge and spin carriers and their precisely controlled injection into quantum dot (QD) nanostructures are a key element of solid-state on chip quantum information networks. The acousto-electric transport by radio frequency surface acoustic waves (SAWs) has been investigated for more than 15 years for one- and two-dimensional electron systems \cite{Talyanskii1997,*Rotter:99a} and dissociated, photogenerated electron-hole ($e$-$h$) pairs \cite{Rocke:97,*Alsina:01,*Kinzel:11}. Recent key breakthroughs based on "surfing" transport include the demonstration of high-fidelity SAW-mediated transfer of single electrons between distant electrostatically defined QDs \cite{Hermelin:11,*McNeil:11} and sequential $e$ and $h$ injection into remote optically active QD nanostructures \cite{Boedefeld:06,*Couto:09,*Hernandez:12} for single photon generation. The latter approach is particularly tantalizing since, in strong contrast to electrostatic QDs, an efficient conversion of a stationary electron or hole encoded spin qubit to a flying photonic qubit for long distance quantum transmission becomes feasible. This process consists of three major steps: (i) carriers being picked up by the SAW, (ii) acousto-electric transport and (iii) injection of the transported carriers into the QD. Out of these three steps, the acousto-electric transport has been studied in greatest detail and, most noteworthy, its spin-preserving nature has been clearly demonstrated in a number of experiments \cite{Sogawa:01a,*Stotz:05,*Hernandez-Minguez2010}. However, to fully exploit the potential of this approach, a fundamental understanding of the onset of the acousto-electric transport and the injection dynamics in the zero-dimensional confined energy levels of the QD is crucially required. Previous optical investigations on SAW control of QD nanostructures mainly focused on dynamic tuning of the optical transitions by the oscillating strain fields \cite{Gell:08,Metcalfe:10} and the control of the carrier injection into single QD emitters by the SAW-induced piezoelectric potentials. However, the latter experiments suffered from several shortcomings, such as studies have been performed on QDs of relatively poor optical quality \cite{Boedefeld:06,*Couto:09,*Hernandez:12} or completely off-resonant carrier generation and time-integrated detection schemes \cite{Voelk:10b,Voelk:11a,Voelk:12} have been used. So far, these limitations severely hampered clear experimental observations and the identification of the underlying physical processes.\\

Here, we report on the time-domain optical observation of acoustically regulated $e$ and $h$ injection into a single optically active semiconductor QD at the onset of acousto-electric transport. For our experiments we use a sample containing self-assembled QDs, which are the established prototype, high quality interface for conversion of single excitons to single photons \cite{Michler:00,*Shields:07}. Spatially resolved, stroboscopic optical spectroscopy allows to deliberately inject carriers at a well-defined local phase of the SAW. Using this unique capability we show, that the time during the acoustic cycle, at which carriers are generated, predetermines both the onset of acousto-electric transport in the semiconductor matrix and the resulting excitonic occupancy state of the QD. This mechanism manifests itself in intensity oscillations between emission lines arising from neutral and charged excitons as the excitation time is tuned over the acoustic cycle. Moreover, the emission of each exciton species exhibits additional characteristic beatings in the time domain. Most remarkably, the oscillations of different exciton species are found to be highly correlated, an observation that can be attributed to conditional time-dependent generation processes. All experimental observations are in excellent agreement with numerical simulations of the acoustically induced charge carrier dynamics in a two-dimensional continuum coupled to the QD. The calculated carrier trajectories clearly visualize the impact of the acoustically programmed electric environment on the onset of acousto-electric transport.\\

\section{Sample structure and stroboscopic, time-integrated optical spectroscopy}
\begin{figure*}[bth]
	\includegraphics[width=1.95\columnwidth]{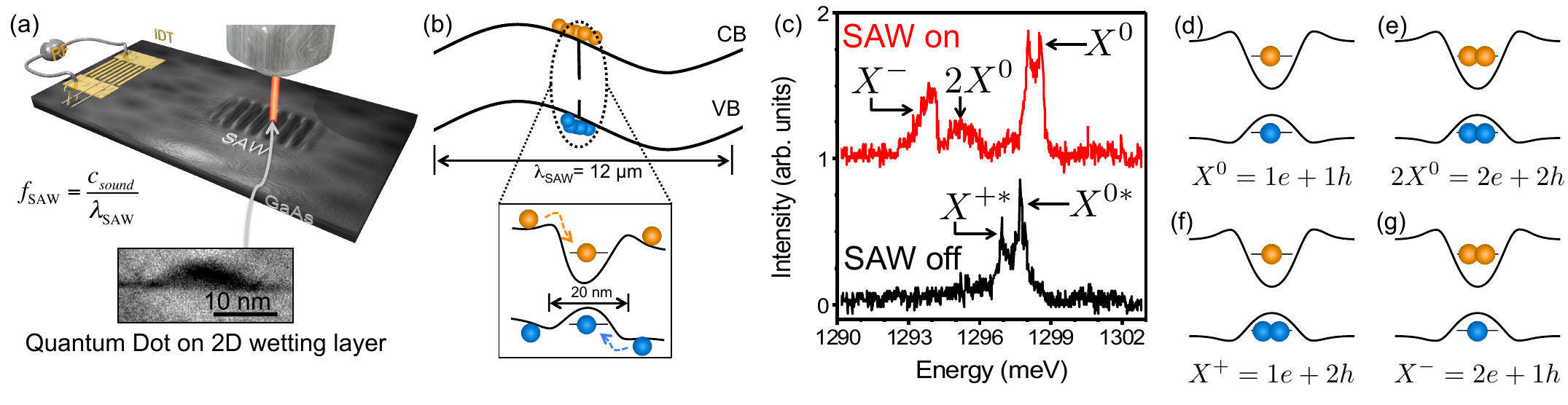}
	\caption{(a) Schematic of our experiment consisting of an IDT generating the SAW which interacts with a single QD. (b) Type-II band edge modulation induced by the SAW. Since the size of the QD is almost three orders of magnitude smaller than the acoustic wavelength, the QD acts as a point-like, optically active deep trap for $e$'s (orange) and $h$'s (blue). (c) Time-integrated emission of a single QD without (black) and with SAW applied (red) and stroboscopic excitation as sketched in (b). In addition to an overall increase of the PL intensity, we observe pronounced intensity exchange between different excitonic complexes confined in the QD. (d-g) Four different s-shell exciton configurations observed in the spectra.}\label{fig1}
\end{figure*}

In our experiments, we studied a sample containing a single layer of self-assembled ${\rm In_{0.5}Ga_{0.5}As/GaAs}$ QDs of ultra-low areal surface density grown by molecular beam epitaxy \cite{Krenner:05a}. These QDs nucleate on top of a wetting layer (WL) which acts as a disordered QW and allows for two-dimensional charge carrier transport. On the sample surface we patterned interdigital transducers (IDTs) which allow for the excitation of Rayleigh-type SAWs of wavelength $\lambda_{\rm SAW}= 12\, {\rm \mu m}$ propagating at the speed of sound, $c_{\rm sound}\sim 2900\,{\rm m/s}$. This wavelength converts to an acoustic frequency of $f_{\rm SAW}=c_{\rm sound}/\lambda_{\rm SAW}=240\,\mathrm{MHz}$ at low temperatures $(T=10\:\mathrm{K})$. The corresponding acoustic period of $T_{\rm SAW}=4.2\,\mathrm{ns}$ ensures full access to the SAW induced dynamics for the given temporal resolution of $\sim 300\,{\rm ps}\sim T_{\rm SAW}/14$ of our setup. As sketched schematically in Fig. \ref{fig1} (a) carriers are photogenerated by $\sim90~{\rm ps}\sim T_{\rm SAW}/47$ pulses of a $850\:\mathrm{nm}$ externally triggered diode laser. The carrier relaxation and radiative recombination times for the self-assembled QDs studied here are $\tau_{relax}\sim 40\,{\rm ps}$ and $\tau_{rad}\sim 1\,{\rm ns}$, respectively \cite{Heitz1997,*Ohnesorge1996,*Wesseli:06,*Moody2011,*Dalgarno2008}. Therefore, when compared to $T_{\rm SAW}/4=1.05\,\mathrm{ns}$, carrier capture occurs quasi-instantaneously while radiative processes occur on time scales set by our acoustic clock. Within the area of the diffraction limited laser focus, charge carriers are generated within the two-dimensional WL continuum at the position of a single QD located in the propagation path of the SAW. The piezoelectric potential of the SAW modulates the conduction band (CB) and valence band (VB) and induces a Type-II band edge modulation depicted schematically in Fig. \ref{fig1} (b). This modulation in turn gives rise to acousto-electric fields which induce drift and diffusion dynamics of the photogenerated carriers within the plane of the WL. In the limit of high acoustic amplitudes electrons $(e$'s$)$ and holes $(h$'s$)$ are conveyed in the plane of the WL by the SAW once they are transferred to their respective stable points in the conduction and valence band. Thus, the length scale of these acoustically induced carrier dynamics in the WL is defined by $\lambda_{\rm SAW}$/2. As sketched in Fig. \ref{fig1} (b), the QD represents a point-like deep trap for $e$'s and $h$'s in this acoustically programmed potential landscape since the dot's dimensions of $\sim 20-30$\,nm are almost three orders of magnitudes smaller than $\lambda_{\rm SAW}$/2. To resolve the SAW-induced carrier dynamics in the WL and the emission characteristics of the QD, we perform stroboscopic micro-photoluminescence (s$\mu$-PL) spectroscopy using established time-integrated \cite{Voelk:11a,Fuhrmann:11} and time-resolved detection. In this excitation scheme, the laser repetition frequency and $f_{\rm SAW}$ are actively locked. By tuning the relative phase of these two oscillations, we photoexcite at a well-defined time during the acoustic cycle as depicted schematically in Fig. \ref{fig1} (b) with $e's$ and $h$'s in orange and blue, respectively. Details of the sample and the experimental techniques can be found in the Appendix. For all experiments presented here, we generated the SAW by applying an rf power of $P_{rf}=+28\,{\rm dBm}$ to the IDT and carriers are photo-generated directly at the QD position. At this high acoustic power level, acousto-electric charge conveyance for \emph{both} carrier species, $e$'s \emph{and} $h$'s, is fully developed. This mechanism gives rise to a dramatic change of the recorded time-integrated optical emission from a single QD which we discuss in the following.\\

In Fig. \ref{fig1} (c) we compare two time-integrated spectra recorded from a single, isolated QD without (black line) and with a SAW applied (red line). In these spectra we identify signatures arising from recombination of the four different s-shell excitons confined in the dot, the neutral exciton $(X^0=1e+1h)$, the positive $(X^+=1e+2h)$ and negative trions $(X^-=2e+1h)$ and the neutral biexciton $(2X^0=2e+2h)$. Schematics of these different configurations are presented in Fig. \ref{fig1} (d-g). With no SAW applied, we observe two main emission lines which we attribute to the neutral exciton  at $E_{X^{0*}}=1297.7\,{\rm meV}$ and the positive trion $(X^{+*}=1e+2h)$. These lines are substantially broadened and shifted due to random electric fields arising from charge noise induced by dynamic trapping and release of photogenerated carriers in the disordered WL potential. This fact is indicated by the superscript $^*$. Houel \textit{et al.} \cite{Houel2012} reported a similar observation for charges trapped at an heterointerface above a QD. Their effect is less pronounced due to the large \emph{vertical} separation of 40\,nm between the QD and the localization site. 
The spectrum with the SAW applied (red) changes dramatically: the dynamic piezoelectric fields induced by the SAW depopulate traps and remove the charges. This in turn reduces charge noise \cite{Voelk:10b} and restores the unperturbed exciton energy which is then spectrally tuned \cite{Gell:08,Metcalfe:10}. This dynamic shift of the transition energy gives rise to the apparent broadening observed in the time-integrated spectrum. In addition to the emission of the $X^0$ at $E_{X^0}=1298.4\,{\rm meV}$, we observe  a clear signal of the negative trion, $X^-$, at $E_{X^-}=1293.7\,{\rm meV}$ and a weaker signature of the neutral biexciton, $2X^0$, at $E_{2X^0}\sim 1295.5\,{\rm meV}$. At this point the assignment of these different emission lines remains based on established, typical renormalization energies for this type of QD \footnote{For QDs grown under these conditions $X^+$ is shifted towards lower energies compared to $X^0$ \cite{Zecherle2010,*Muller2013}.} \cite{Zecherle2010,*Muller2013} and will be further confirmed by our time-resolved spectroscopy and numerical modelling.

\begin{figure*}[htb]
	\includegraphics[width=1.95\columnwidth]{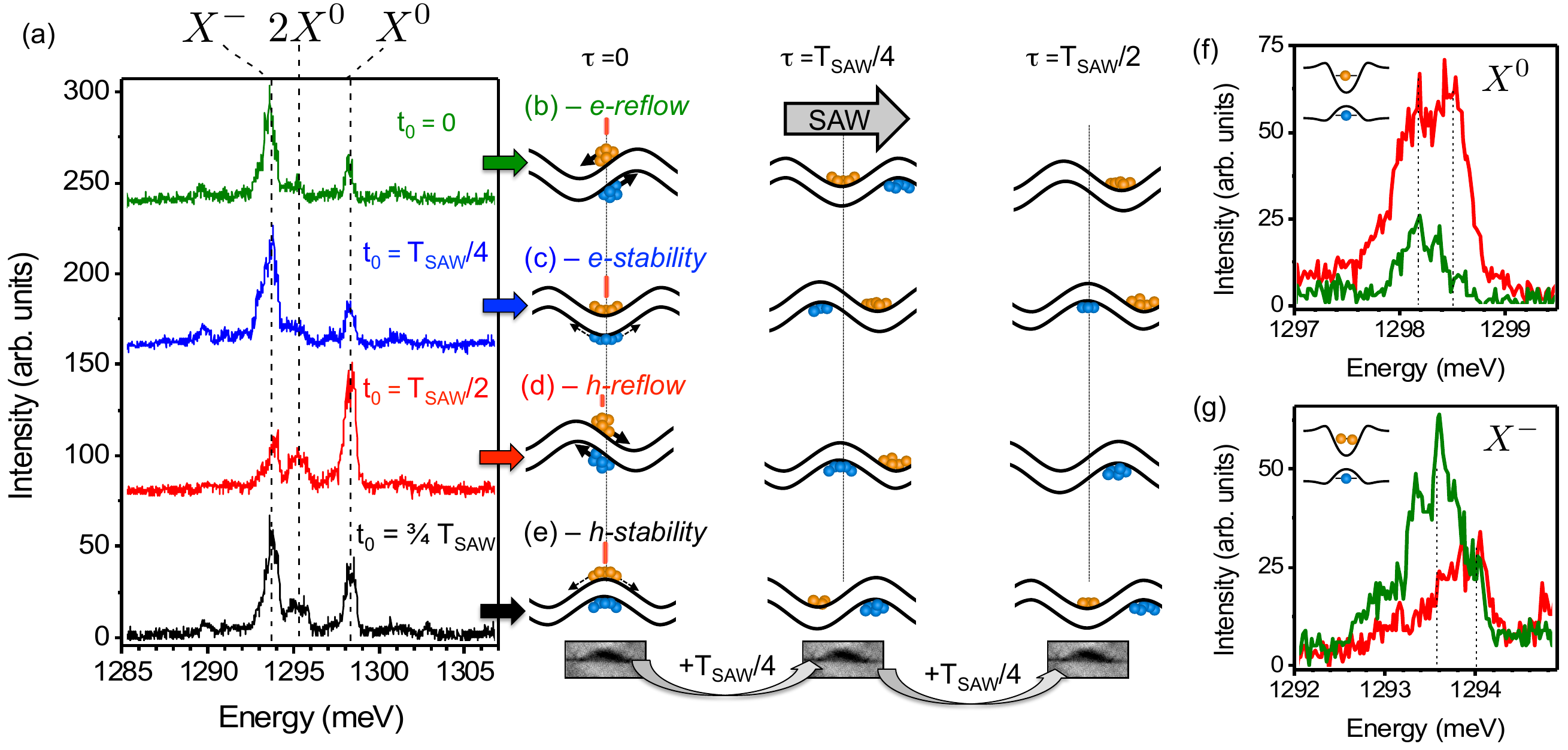}
	\caption{Stroboscopic, time-integrated PL spectra of a single QD (a) when optically generating carriers at four characteristic times during the acoustic cycle. (b-e) Schematic of the corresponding excitation conditions and motion of $e$'s (orange) and $h$'s (blue) in the Type-II band edge modulation. The center and right panels show the time evolution in steps of $T_{\rm SAW}/4$ for a SAW propagating from left to right. (f+g) $X^0$ and $X^-$ emission lines at \emph{e-} and \emph{h-reflow}, respectively. Clear spectral shifts due to dynamic acoustic tuning are resolved for both lines. The dashed lines in (f+g) indicate a spectral tuning range of $\sim 0.4{\, \rm meV}$. Line colors correspond to the excitation conditions (b+d).}\label{fig2}
\end{figure*}

In Fig. \ref{fig2} (a) we show time-integrated s$\mu$-PL spectra which we recorded from a single QD  with laser excitation at four distinct times during the acoustic cycle. 
We calibrated the time of laser excitation by probing the carrier dynamics in the WL as described in Ref. \cite{Schuelein:12a}. 
In the detected QD signal, clear correlated intensity oscillations between the neutral exciton $(X^0)$  and negative trion $(X^-)$ are resolved as we tune the time of excitation over the acoustic cycle. In contrast to our previous studies \cite{Voelk:10b} the positive trion $(X^+)$ is largely suppressed \footnote{We attribute these key differences to the clean quasi-resonant pumping of the WL. Under these conditions, carriers are generated in a two-dimensional quantum well with confinement in the plane of SAW propagation. Therefore, no additional carrier dynamics in the vertical direction occur unlike in the case of off-resonant pumping in the bulk GaAs matrix.} and $X^-$ is the dominant charged exciton species formed. Furthermore, the contrast of the intensity oscillation is significantly enhanced compared to the one observed for columnar quantum posts \cite{Voelk:11a} and weak spectral broadening of up to $\sim 0.4\,{\rm meV}$ due to dynamic SAW-tuning \cite{Gell:08,Metcalfe:10} is observed [cf. Fig. \ref{fig2} (f+g)].\\

We define the time $\tau$ as the temporal delay \emph{after} the optical pump at $t_0$. The resulting excitation conditions and the time evolution of the acoustically induced potential for a SAW propagating from left to right are shown schematically in Fig. \ref{fig2} (b-e). We refer to excitation conditions (b), $t_0=0$ and (d) $t_0 = T_{\rm SAW}/2$ as \emph{e-reflow} and \emph{h-reflow}, respectively. Here carriers are generated at maximum electric field and either $e$'s or $h$'s initially drift against the SAW propagation direction and are transported back to the QD position after $\tau \sim T_{\rm SAW}/4$. The opposite carrier species drifts with the SAW and continuously moves away from the QD position. The excitation conditions (c) and (e) are referred to as \emph{e-stability}, $t_0 = T_{\rm SAW}/4$, and \emph{h-stability}, $t_0 = 3T_{\rm SAW}/4$, under which the respective carriers are generated at their stable point in the CB and VB. The opposite charge carrier species is generated at the unstable point. After transfer to their corresponding two adjacent stable points, these carriers arrive at the QD position after $\tau\sim T_{\rm SAW}/2$. These acoustic charge conveyance processes are shown in the center and right column for two $T_{\rm SAW}/4$ steps after the optical pump. Using these definitions, we summarize the observed intensity oscillations as follows: Emission from $X^-$ dominates the spectrum for \emph{e-reflow} and \emph{e-stability}. In contrast, dominant $X^0$ emission is observed for \emph{h-reflow} and $X^0$ and $X^-$ exhibit similar intensities for \emph{h-stability}. 
\\
The key observations for three of the four excitation conditions can be qualitatively understood. For $e$-  and \emph{h-reflow} [cf. green and red spectra in Fig. \ref{fig2} (a), respectively], carriers are generated in a region of maximum electric field. $X^-$- and $X^0$-formation occur either quasi-instantaneously  on a ps-timescale with the optical pump or time delayed when $e$'s and $h$'s are being transported back to the QD position by the propagating SAW. At \emph{e-stability}, the pump laser generates $e$'s at a field-free, stable CB minimum. This locally confines these carriers and inhibits fast initial dynamics. This in turn leads to preferential $e$ capture and $X^-$-formation. Similar reasoning cannot be applied to explain the pronounced emission of $X^-$ and weak emission of $2X^0$ for \emph{h-stability}. Moreover, the detailed time dependence and different formation processes cannot be resolved in such a stroboscopic but time-integrated experiment. One clear indication for time-dependent carrier injection is the observation of a pronounced energy broadening of $0.4\,{\rm meV}$ for the $X^0$ emission for \emph{h-reflow} (red) which is absent for \emph{e-reflow} (green). The corresponding spectra are compared in Fig. \ref{fig2} (f) and the apparent broadening is marked by dashed vertical lines. Since the initial local strain fields at $\tau =0$ at the QD position are equivalent at these two local acoustic phases, the emission detected for \emph{h-reflow} has to stem in parts from $X^0$ formed at later times $\tau\gg0$. Then the QD is under a different strain bias \cite{Gell:08,Metcalfe:10} or different lateral electric field \cite{Vogel2007,*Kaniber:11a}, which both give rise to a shift of the emission energy. In addition, signatures of dynamic spectral tuning are also observed for $X^-$ when comparing the spectra under \emph{h-reflow} (red) and \emph{e-reflow} (green) in Fig. \ref{fig2} (g).\\ 

\section{Time-resolved spectroscopy and comparison to numerical simulations of charge carrier trajectories}

\begin{figure*}[thb]
	\includegraphics[width=1.95\columnwidth]{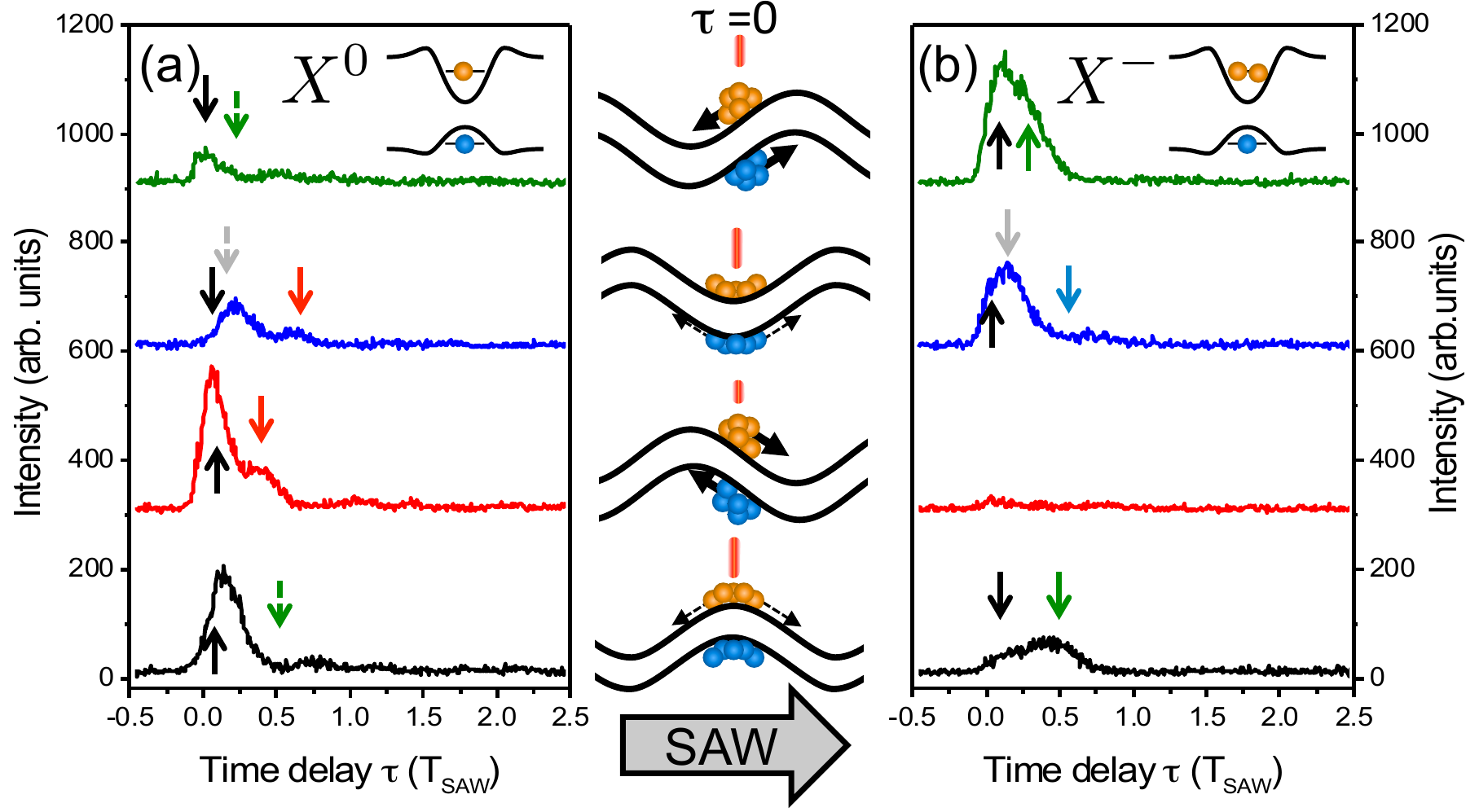}
	\caption{Time-resolved PL of the $X^0$ (a) and $X^-$ emission (b) recorded at the four different stroboscopic excitation conditions shown in the center. Black and grey arrows mark formation and recombination with the optical pump. The color code of arrows marking time-delayed formation processes is the same as in Figs. \ref{fig4} and \ref{fig5}. Dashed and solid color arrows in panel (a) and (b) mark suppression of $X^0$ and enhancement of $X^-$ emission due to conditional $X^0\rightarrow X^-$ formation.}\label{fig3}
\end{figure*}

To fully resolve the dynamic acoustic preparation of the different occupancy states, we performed time-resolved spectroscopy. In Fig. \ref{fig3} we present PL transients recorded from the $X^0$ and $X^-$ emissions in panels (a) and (b), respectively \footnote{The low count rates in our pulsed SAW scheme did not allow for time-resolved experiments on $2X^0$.}. For each exciton species, these transients were recorded at the four distinct, previously defined, stroboscopic excitation conditions which are depicted schematically in the center part of Fig. \ref{fig3}. The PL intensities measured from $X^0$ and $X^-$ are plotted as a function of time delay $\tau$ after laser excitation ($\tau =0)$ in units of $T_{\rm SAW}=4.2\,{\rm ns}$. Clearly, all eight transients do not show simple exponential decays but exhibit non-trivial pronounced beatings at up to $\tau \sim 0.8 T_{\rm SAW}$ after the optical pump. These time-delayed emission events at different excitation conditions follow the corresponding acoustic phase shifts. This can be nicely seen most pronounced for the features marked by red arrows in the $X^0$-transients at \emph{e-stability} and \emph{h-reflow}. Furthermore, enhancements of the $X^-$ emission and suppression of that of $X^0$ are highly correlated as indicated by the dashed and solid arrows of the same color in for $X^0$ and $X^-$, respectively. In the following we will show that all characteristic signatures resolved in the PL transients marked by arrows in Fig. \ref{fig3} can be readily understood by the underlying acoustically induced charge carrier dynamics and acousto-electric transport and subsequent conditional carrier capture pathways into the QD's confined energy states.\\
\begin{figure}[htb]
	\includegraphics[width=.95\columnwidth]{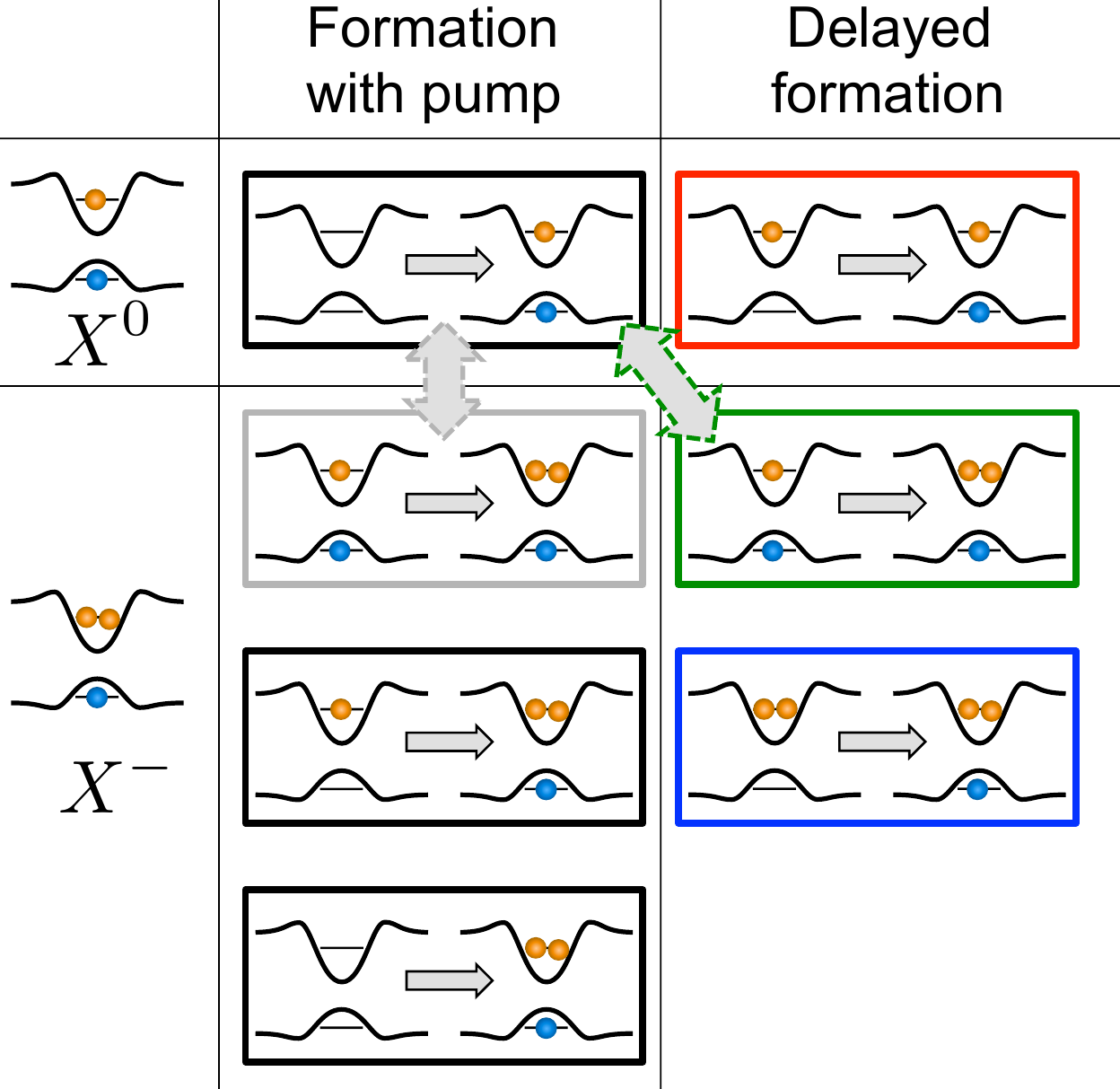}
	\caption{Overview of different formation pathways for $X^0$ and $X^-$ at the time of laser excitation (formation with pump -- black/grey) or after a temporal delay (color coded). Conditional formation pathways $X^0\rightarrow X^-$ are highlighted by the dashed arrows.}\label{fig4}
\end{figure}

We start by considering the main formation pathways for $X^0$ and $X^-$ being summarized in Fig. \ref{fig4}. In general, both species can be formed either during the initial optical pump (left column, marked in black and grey) or time-delayed by acousto-electric charge conveyance (right column, color coded). The two formation conditions differ fundamentally: For formation with the optical pump, \emph{both} carrier species can be injected while for a delayed formation only the \emph{specific}, acoustically transported, carrier species is captured. Since in our experimental data the signal of the \emph{positive} trion $(X^+)$ is largely suppressed, we neglect formation pathways starting with the QD occupied by one or two $h$'s. Applying this limitation, $X^0$ can be formed with the pump by capture of one $e$-$h$-pair or delayed by capture of a conveyed $h$ by the QD containing already a single $e$ (marked in red). In contrast, there exist five formation pathways to create a $X^-$ in the QD. There exist two quasi-instantaneous formation pathways and one delayed from an optically inactive state: the capture of one $e$-$h$-pair into the QD containing a single $e$, the capture of $X^-=2e+1h$ into the empty QD with the optical pump (both marked in black) and the delayed injection of one $h$ into the QD being occupied by $2e$ (blue). In addition, $X^-$ can be formed from the optically active $X^0$ by capture of a single $e$. This injection can occur both with the optical pump (grey) or time-delayed triggered by the SAW (green). Since this $X^0\rightarrow X^-$ process converts a neutral exciton to a trion, it is expected to result in a time-correlated suppression of the $X^0$ emission [dashed arrows in Fig. \ref{fig3} (a)] and an enhancement of the $X^-$ signal [solid arrows Fig. \ref{fig3} (b)]. \\

\begin{figure*}[hbt]
	\includegraphics[width=1.25\columnwidth]{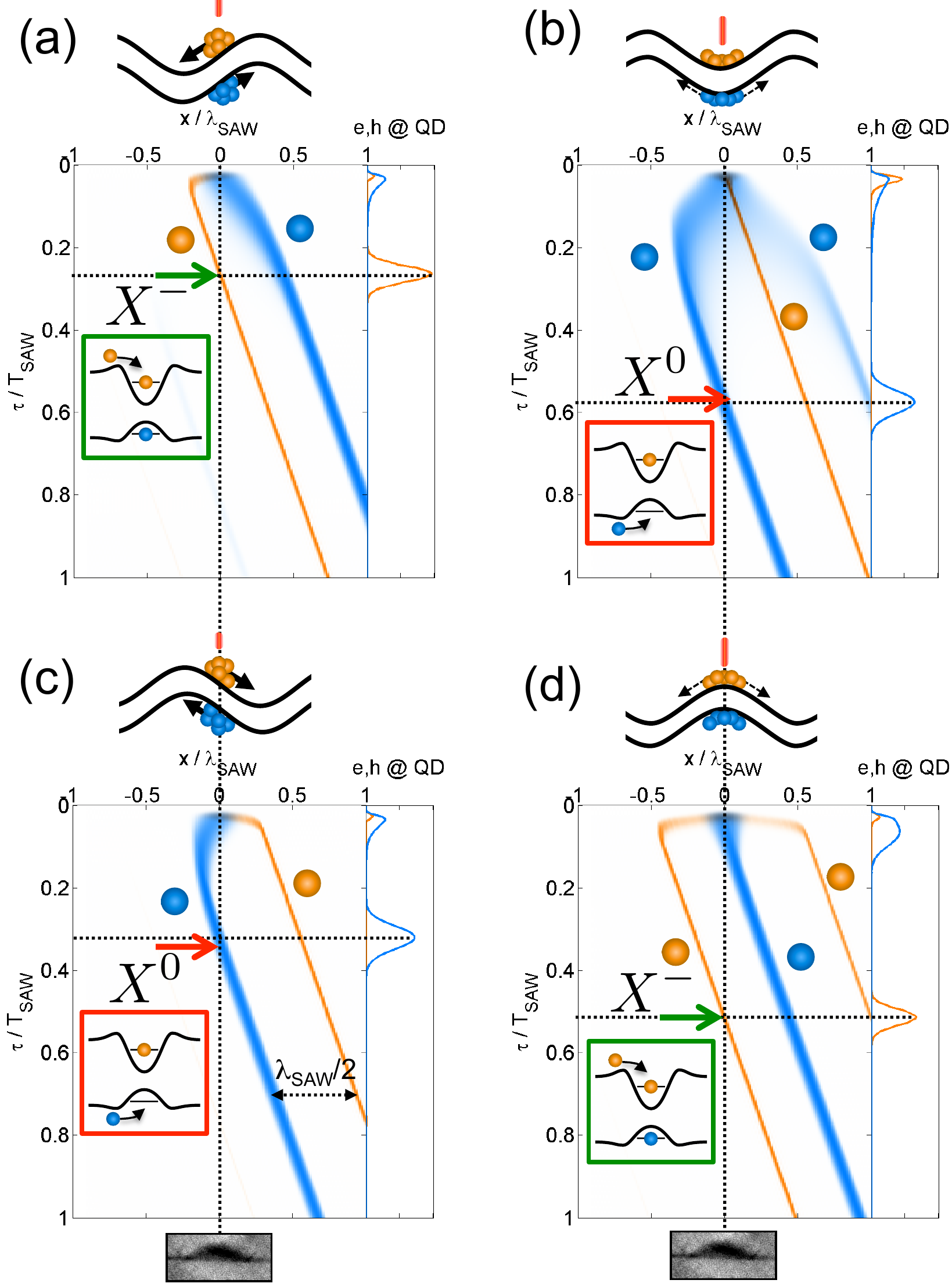}
	\caption{Calculated $e$ (orange) and $h$ (blue) dynamics for the four characteristic excitation conditions (a-d) with the SAW propagating along the horizontal axis from left to right and time increasing from top to bottom. The position of the QD is in the center of each main panel. A cross section of the $e$- and $h$-densities at the position of the QD is plotted in the righthand side panels. Horizontal dashed lines mark SAW-controlled, time-delayed arrival of carriers at the QD positions and arrows mark the experimentally observed signatures.}\label{fig5}
\end{figure*}

All time-delayed formation pathways depend in a non-trivial manner on the precise time of arrival of the acoustically transported carriers. We calculate the underlying SAW-induced carrier dynamics in the WL at the four different excitation conditions: We numerically solve the semi-classical drift and diffusion equations for $e$'s and $h$'s from which we obtain the trajectories for the densities of both carrier species as a function of time along the SAW propagation direction \cite{Garcia2004,Schuelein:12a}. The calculated trajectories for the $e$- and $h$-densities are plotted in Fig. \ref{fig5} for \emph{e,h-reflow} (a+c) and \emph{e,h-stability} (b+d), respectively. Time ($\tau$ in units of $T_{\rm SAW}$) increases on the vertical axis from top to bottom and the SAW propagates along the horizontal axis ($x$ in units of $\lambda_{\rm SAW}$) from left to right. In these plots, the $e$-density is encoded in the orange color code, while the $h$-density is encoded in blue. Vertical cuts of the two carrier densities at the position of the QD $(x=0)$ are evaluated in the righthand side panel of each plot. The calculated $e$- and $h$-density evolutions exhibit for all four excitation conditions characteristic fingerprints which we can directly relate to our experimental data.\\

We start this comparison by evaluating the carrier injection with the optical pump: (i) For all four scenarios both carrier species are acoustically conveyed at $c_{\rm sound}$ in their corresponding stable points of the SAW-induced potential. This manifests itself in a constant slope and a constant spatial separation of $\lambda_{\rm SAW}/2$. (ii) For excitation of carriers at $e$- or \emph{h-reflow}, the piezoelectric field rapidly dissociates the $e$-$h$-pair and transfers each species to its stable point in the bandstructure. Since the initial dissociation and transfer is electric field induced, initial capture can only occur at short times after the pump. Therefore, we expect a fast rise time of the PL emission which is also observed and marked by black arrows in the corresponding (green, red) experimental transients in Fig. \ref{fig3}. (iii) In contrast, at $e$- or \emph{h-stability}, the carrier species generated at an unstable point transfer to the \emph{two} neighboring stable points. This transfer is initially driven by carrier drift-diffusion, limited by the transferred carriers' mobility. Therefore, this process is significantly slower for $h$'s compared to $e$'s due to the $h$'s lower mobility, which is reflected by the pronounced and weak spatial spreading of the calculated $h$- and $e$-distributions up to $\tau\sim 0.5T_{\rm SAW}$ in Fig. \ref{fig5} (b) and (d), respectively. In the experimental data for $X^0$ in Fig. \ref{fig3} (a), we observe slow (intermediate) rise time at \emph{e-stability} (\emph{h-stability}) as marked by the black arrows. The dominant contribution to the pronounced shift of the maximum at \emph{e-stability} to $\tau \sim0.25 T_{\rm SAW}$ arises from a conversion of $X^0\rightarrow X^-$. This underlying process of capture of an additional $e$ is enhanced due to the comparably high density of this carrier species under these excitation conditions. This interpretation is further confirmed by the observed temporal spreading of the $X^-$ at short times after laser excitation. Moreover, the maximum of the $X^0$ emission occurs with a finite time delay after the maximum of $X^-$. To emphasize this delay, the time of the $X^-$ maximum is marked by the dashed and solid grey arrows in Fig. \ref{fig3} (a) and (b), respectively. At this time the process $X^0\rightarrow X^-$ breaks down since $e$'s have been removed from the QD position by the SAW. \\

In addition to processes occurring at short timescales after optical excitation, $e$'s and $h$'s transferred to a stable point in the SAW potential can be acoustically conveyed and injected into the QD at later times. Delayed emission events reflect the time of arrival of the carriers at the QD position and monitor the onset of acousto-electric charge conveyance. Our simulations predict the arrival of conveyed holes under \emph{e-stability} and \emph{h-reflow} at $\tau\sim0.58 T_{\rm SAW}$ and $\tau\sim0.32 T_{\rm SAW}$, respectively. In the experimental data we resolve the corresponding signatures marked by red arrows in the $X^0$ PL-transients [Fig. \ref{fig3} (a)] at precisely these calculated times. The underlying process highlighted in the same color code in Fig. \ref{fig4} is the injection of a single $h$ into the QD being already populated by a single $e$. We want to note that these calculated and measured arrival times are shifted from the trivially assumed values of $\tau=0.5 T_{\rm SAW}$ and $\tau=0.25 T_{\rm SAW}$. This delay arises from the fact that acousto-electric transport most effectively occurs for the carrier drift velocity along the propagation direction of the SAW being equal to the SAW propagation speed. This condition is fulfilled for $v_{e/h}=\mu_{e/h}E_{\rm SAW,\parallel}=c_{sound}$, with $\mu_{e/h}$ and $E_{\rm SAW,\parallel}$ being the carrier mobility and in-plane electric field induced by the SAW, respectively. Since $E_{\rm SAW,\parallel}=0$ at the maximum/minimum of the electric potential, the transport position of $e$'s and $h$'s is shifted slightly in the direction opposite to the SAW propagation. In contrast to the delayed $X^0$ formation, similar signatures stemming from $X^-$ formation from a $2e$ state are not detected in the $X^-$ time transients. The blue arrow in Fig. \ref{fig3} (b) marks the time at which we predict this process from our calculations. The absence of this feature indicates that the probability for $e$ ($2e$)-capture into a QD containing already a single (no) $e$ are highly unlikely. Under \emph{e-reflow} and \emph{h-stability}, we expect the arrival of conveyed $e$'s at the QD position at $\tau\sim0.28 T_{\rm SAW}$ and $\tau\sim0.51 T_{\rm SAW}$, respectively. This shift from the stable conveyance point is significantly smaller than that for $h$'s due to higher mobility for $e$'s. The time-delayed injection of a single $e$ reflects itself in a delayed formation of $X^-$ starting from $X^0$, highlighted in green in Fig. \ref{fig4}. This formation process is observed nicely in the PL transients in Fig. \ref{fig3}: The maximum of the $X^-$ emission is shifted to $\tau\sim 0.5T_{\rm SAW}$ (solid green arrow) at \emph{h-stability}. At exactly the same time we observe a minimum in the $X^0$ transient (dashed green arrow) providing direct evidence of a direct correlation. In the corresponding transients recorded at \emph{e-reflow}, a clear shoulder and a minimum  at $\tau\sim 0.28T_{\rm SAW}$ (green arrow) are again resolved for $X^-$ and $X^0$, respectively. Finally, we compare our experimental data and simulations by marking the experimental signatures by arrows in Fig. \ref{fig5}. We apply the same color codes for Figs. \ref{fig3} and \ref{fig4} to highlight the excellent agreement between the experimental data (arrows) and the calculated times of arrival (dashed lines) which clearly demonstrates the direct correlation between the observed and calculated acoustically induced charge dynamics and the different exciton formation pathways.\\

\section{Conclusions and perspectives}

Taken together, our combined experimental and theoretical results directly resolve the onset of acousto-electric charge conveyance in a two-dimensional system and reveal the different, conditional and correlated formation pathways for different occupancy states of a single QD. The trajectories of $e$'s and $h$'s in the acoustically generated dynamic potential predetermine the time at which these carriers are injected into the QD and, thus, the formed exciton species. Comparing time-domain experiments and computed trajectories, we identify different time-dependent uncorrelated and highly correlated formation pathways for neutral and charged excitons. These, in combination with the adjustable time of carrier excitation, allow for programming of both the emission energy and time. Finally, we comment on two implications and perspectives of our results:\\ \textit{Quantum light sources --} The precisely regulated carrier injection can be readily extended to fully remote carrier injection as outlined in the original proposal by Wiele \textit{et al.} \cite{Wiele:98}. This approach promises a better definition of the resulting occupancy state due to the inherently sequential injection with the order determined by the time of excitation. The emission rate in our experiments was limited by the radiative decay of the QD in an isotropic optical medium, this figure of merit can be optimized using Bragg microcavities. Such type of microcavity is fully compatible with SAW technology \cite{Lima:05} and allows for acousto-electric charge transport due to its planar geometry. To achieve high Purcell-enhancements, high quality oxide-aperture optical nanocavities \cite{Stoltz:05,*Strauf:07} are well suited to push emission rates to the gigahertz regime and eliminate parasitic timing-jitter which is detrimental e.g. for two-photon interference experiments \cite{Patel2010,*Flagg2010}. Our approach could be of particular interest for the \emph{deterministic} preparation of the neutral biexciton state for photo-excitation at \emph{h-reflow} for which the formation of neutral excitons is favored. In this context, our acoustic technique represents a promising alternative to Coulomb blockade based approaches for the preparation of the initial state of the $2X^0\rightarrow 1X^0$ cascade \cite{Benson:00} since $e$ and $h$-injection is strictly sequential. The crucial elimination of the fine-structure splitting, which is required for a high-fidelity source of polarization entangled photon pairs, can be achieved in a similar approach pioneered by Trotta \emph{et al.} \cite{Trotta2012,*Trotta2012a}. For a SAW-based device, static uniaxial strain and the gyrating strain of the SAW could serve as the two independent tuning parameters. For our experiments presented here, we used prototype self-assembled QDs. Here, acousto-electric charge transport occurs in the WL, which exhibits  relatively poor carrier mobilities which in turn require high SAW amplitudes. Thus, further improvements could be achieved by advanced QD architectures providing high mobility transport channels.
In addition to hybrid coupled QD-QW structures \cite{Mazur2010a}, these include self-assembled quantum posts \cite{Li2007,*He:07,*Krenner:09} for which single photon emission \cite{Krenner:08a}, enhanced acousto-electric transport \cite{Voelk:10b} and remote SAW-driven carrier injection \cite{Voelk:12} has been recently demonstrated. For short wavelength operation, "inverted" GaAs/AlGaAs QDs with tunable WL thickness \cite{Rastelli:04a,*Wang:09a} could be the system of choice.\\
\textit{Spin transport and spin-photon conversion:} The discrete emission spectrum and the polarization selection rules allow for a direct spin-photon conversion. Moreover, it provides an alternative read-out scheme for spin qubits localized in a QD or QD molecule \cite{Gywat:09:book,*Atature:06,*Press:08,*Brunner:09,*DeGreve2011,*Greilich2011,*Muller2012} by injecting a remotely generated, unpolarized carrier of opposite charge by a SAW to induce radiative recombination. Since the time of arrival can be precisely triggered and delayed by a SAW, such a read-out scheme would not be limited by fast radiative decay or tunnel processes.
The spin-preserving nature of the underlying acousto-electric transport \cite{Sogawa:01a,*Stotz:05} and the capture of carriers into the QDs \cite{Trumm:05} provide further tantalizing perspectives to extend our approach to acoustic transfer of single spins to an optically active QD for reconversion to the optical domain. Sanada \textit{et al.} recently demonstrated electron spin resonance of SAW conveyed $e'$s without external magnetic fields by controlling the spin-orbit interaction along the acoustic transport channel  \cite{Sanada2013}. Thus, these three parts could be merged to a single spin, manipulation and an optical readout protocol.\\

\begin{acknowledgements}
We gratefully acknowledge financial support by the Deutsche Forschungsgemeinschaft (DFG) via Sonderforschungsbereich SFB 631, the Emmy Noether Program (KR3790/2-1) and the Cluster of Excellence {\it Nanosystems Initiative Munich} (NIM).
\end{acknowledgements}

\appendix

\section{Experimental details}
\subsection{Sample}
Our sample contained a single layer of self-assembled ${\rm In_{0.5}Ga_{0.5}As/GaAs}$ QDs (nominal coverage $4\,\mathrm{ML}$) which is located 280\,nm below the sample surface. During deposition of the ${\rm In_{0.5}Ga_{0.5}As}$ layer, we intentionally stopped the substrate rotation to realize a gradient of the In coverage across the substrate. This material gradient results in the formation of self-assembled QDs in regions of high coverage and a thin wetting layer (WL) without QDs in the low coverage region \cite{Krenner:05a}. In the transition region of low QD areal density, in-plane QD separations exceed $1{\,\rm\mu m}$ which in turn allows for optical addressing of individual QDs without the need of masking techniques. We fabricated a set of Ti/Al interdigital transducers (IDTs) with a periodicity of $2p=\lambda_{\rm SAW}= 12\, {\rm \mu m}$ which allowed for the excitation of a $f_{\rm SAW}=240\,\mathrm{MHz}$ SAW directly on the GaAs surface.
\subsection{Optical spectroscopy}
All optical experiments were performed at $T=10\:\mathrm{K}$ in a He-flow cryostat in a conventional $\mu$-photoluminescence (PL) setup. For the optical excitation, we use an externally triggered pulsed laser diode emitting $\sim90~{\rm ps}$ long pulses at $850\:\mathrm{nm}$. The laser beam was focused onto the sample by a $50\times$ objective to a $\sim2\:\mu\mathrm{m}$ diameter spot. We employed low excitation laser power of 250-400\,nW to ensure preferential generation of single exciton species $X^0$ and $X^-$ and to avoid screening of the piezoelectric potentials by photogenerated carriers. The emitted PL was collected by the same lens and  dispersed by a $0.5\:\mathrm{m}$ single grating monochromator. Time-integrated spectra were recorded by a multi-channel $l{\rm N_2}$-cooled silicon CCD detector. For time-resolved experiments we performed time-correlated single photon counting (TCSPC) using a single-channel silicon single photon counting module with a temporal resolution of $\sim 300\,{\rm ps}$. This time resolution corresponds to $\sim0.07 T_{\rm SAW}$, which  defines the experimental error of our time-resolved optical spectroscopy.\\
\subsection{Synchronization of pulsed laser and SAW}
To resolve the full SAW-driven dynamics of the PL, we employed an established, phase-locked, stroboscopic excitation scheme \cite{Voelk:11a,Fuhrmann:11,Schuelein:12a} for which we actively synchronized a rf signal generator for SAW excitation and a pulse generator to trigger the laser diode by setting $n\cdot f_{{\rm laser}}=f_{{\rm SAW}}$ with $n$ integer. This technique enables us to tune the time delay $t_0$ between the laser excitation and the SAW over two full acoustic cycles $-T_{\rm SAW}\leq t_0\leq+T_{\rm SAW}$ by tuning the \emph{relative} phase $-360^{\circ}\leq\varphi\leq+360^{\circ}$ of the rf signal which generates the SAW. For all experiments we applied SAW pulses (duration $t_{\rm pulse}=1\,{\rm\mu s}$, repetition frequency $f_{\rm rep, SAW}=100\,{\rm kHz}$) to avoid spurious sample heating effects and interference with counter-propagating, reflected SAW pulses.

\section{Numerical simulations}

The calculation of the charge carrier dynamics in the wetting layer is restricted to one spatial dimension and is based on the drift and diffusion of electrons and holes. Two coupled differential equations describe the time evolution of the two charge carrier densities ${n}$ (electrons) and ${p}$ (holes) \cite{Garcia2004,Schuelein:12a}:
\begin{equation}
   \begin{aligned}
   \frac{\partial n} {\partial t} &= D_{n} \cdot \frac{\partial^2 n}{\partial x^2} \:{+}\:  \mu_{n} \cdot \frac{\partial (E_{x} \cdot n)}{\partial x} + G(x,t) - R(n,p) \\
   \frac{\partial p} {\partial t} &= D_{p} \cdot \frac{\partial^2 p}{\partial x^2} \:{-}\:  \mu_{p} \cdot \frac{\partial (E_{x} \cdot p)}{\partial x} + G(x,t) - R(n,p)
   \end{aligned}
\end{equation}
The generation rate $G$ corresponds to the pulsed laser excitation: We excite $\sim10$ pairs of electrons and holes during a $50\,{\rm ps}$ long laser pulse in a $1.5\,{\rm \mu m}$ diameter profile. The temporal and spatial profile are assumed to be Gaussian. The carrier recombination rate $R$ is proportional to the exciton generation rate $n\cdot p$. The exciton population is assumed to be a mono-exponential radiative decay with a lifetime of $600\,{\rm ps}$. The in-plane electric field, $E_{x}$, has two contributions: the locally induced electric field by the net charge density $(p-n)$ and the superimposed contribution of the SAW. The latter is given by the local gradient of the acoustically induced piezoelectric potential. 
Furthermore, we use mobilites $\mu_n = 0.01\,{\rm \frac{m^2}{Vs}}$ and $\mu_p = 0.001\,{\rm \frac{m^2}{Vs}}$ for electrons and holes, respectively. $D_{n,p}=\mu_{n,p} k_BT/e$ denote the $e$ and $h$ diffusion coefficients. Here $k_B$ is the Boltzmann constant, $T=10\,{\rm K}$ the temperature and $e$ the elementary charge.

\end{document}